# Spectral maximum in photoconductance of a quantum point contact

V.L. Alperovich[1,2] *, D.M. Kazantsev[1,2], V.A. Tkachenko[1,2]

[1] Rzhanov Institute of Semiconductor Physics, 630090, Novosibirsk, Russia

[2] Novosibirsk State University, 630090, Novosibirsk, Russia

**Abstract**

A counter-intuitive disappearance of the giant terahertz photoconductance of a quantum point contact (QPC) under increase in the photon energy, which was discovered experimentally (Otteneder et al., Phys. Rev. Applied 10 (2018) 014015) and studied by the numerical calculations of the photon-stimulated transport (O.A. Tkachenko et al., JETP Lett. 108 (2018) 396), is explained here by using qualitative considerations about the momentum conservation upon absorption of terahertz photons. The spectra of photon-stimulated transmission through a smooth one-dimensional barrier are calculated on the basis of the perturbation theory. These calculations also predict the spectral maxima for optical transitions from the Fermi level to the top of the potential barrier. Within the proposed physical picture, the widths of the spectral maxima are estimated, and the evolution of the shape of the spectra with a change in the position of the Fermi level is qualitatively explained.

* alper@isp.nsc.ru

Diverse physical phenomena caused by the influence of ac electromagnetic fields on the quantum transport of electrons in solid-state nanostructures have been widely studied for the last several decades [1,2,3,4,5,6,7,8,9,10,11]. While for the superconductor systems and semiconductor structures with abrupt potential jumps and bound or quasi-bound electronic states these phenomena were studied both experimentally and theoretically [1,2,3,4,5,6], for the semiconductor quantum point contacts only theoretical investigations [7,8,9] were available until recently.

The authors of [12,13,14] have discovered experimentally and studied numerically the effect of the giant photoconductance of a quantum point contact (QPC). The effect occurred in the tunneling mode, under irradiation by terahertz radiation with photon energy $\hbar\omega_0 = 2.85$ meV, close to the difference between the top of the potential barrier and the Fermi energy $\hbar\omega_0 \approx U_0 - E_F$ (Fig.1a). The effect was explained by the photon-stimulated transport (PST) of electrons due to the absorption of photons. However, the disappearance of the photoconductance observed in [12] for a higher photon energy $\hbar\omega_1 = 6.74$ meV, has not received a clear qualitative explanation, although it agrees with the results of the numerical calculations [12,13]. In this paper, we propose such an explanation based on semiclassical considerations of the momentum conservation at PST.

Under the action of an electromagnetic wave with a frequency $\omega$, the energy conservation law allows electrons to undergo transitions to the Floquet states with energies $E_0 \pm n\hbar\omega$, where $E_0$ is the initial energy, $n = 1, 2,…$ corresponds to the number of absorbed (+) or emitted (-) radiation quanta. Due to the momentum conservation law, the absorption of photons should occur with simultaneous scattering in momentum: by phonons or impurities in the bulk of the crystal, or by interacting with nanostructures. For nanostructures with quantum levels or quasi-levels, PST resonances usually appear as photon replicas of resonances in electron transport in the absence of radiation. In a QPC with a single smooth barrier, in which there are no abrupt potential jumps, levels, or quasi-levels, such replicas cannot be observed. Nevertheless, in the photoconductance of such a QPC, a spectral resonance can be observed under the photon-induced electron transitions to the top of the barrier [7,13].

The reason for the resonance is illustrated in the energy-momentum diagram (Fig. 1b), which shows the electron dispersion laws near the stopping point and near the top of the barrier, as well as optical transitions with photon energies $\hbar\omega_0$ и $\hbar\omega_1$. It can be seen that for the "resonant" photon energy $\hbar\omega_0 \approx U_0 - E_F$, the optical transition from the bottom of the lower parabola to the bottom of the upper parabola is vertical and does not require additional scattering in momentum; therefore, the probability of such a transition is high. On the contrary, for $\hbar\omega_1 > \hbar\omega_0$, the transition to a state with a high kinetic energy of an electron over the top of the barrier requires simultaneous scattering in momentum (shown by the dashed line in Fig. 1b), so the probability of such a transition is small due to the small probability of acquiring a large momentum under transfer through a smooth barrier. Therefore, for $\hbar\omega > \hbar\omega_0$, the effect of PST decreases with the increase in the photon energy due to the decrease in the photon absorption probability. For $\hbar\omega < \hbar\omega_0$, i.e., when the electron final energy is below the top of the barrier, PST increases with increasing photon energy due to the increase in the electron tunneling probability through the barrier. As a result, the magnitude of PST reaches its maximum at $\hbar\omega \approx \hbar\omega_0$. A similar picture of the formation of spectral maxima is also valid for multiphoton transitions with $n > 1$.

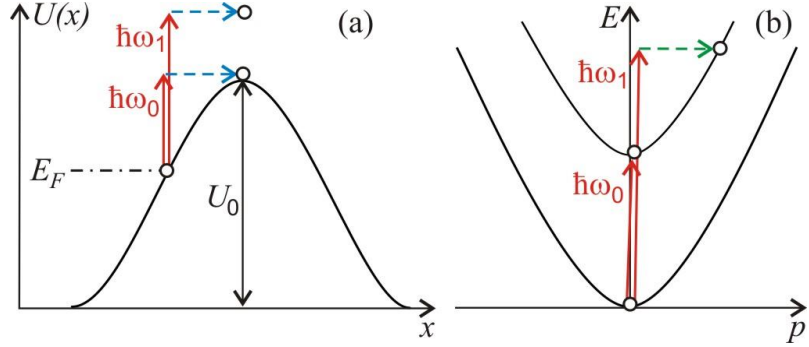

Fig.1. Illustration of the PST through a smooth potential barrier in the energy-coordinate (a) and energy-quasimomentum (b) diagrams when electrons absorb photons with energy $\hbar\omega_0 \approx U_0 - E_F$, corresponding to the transition to the top of the barrier, and for a higher photon energy $\hbar\omega_1 > \hbar\omega_0$. In fig. 1 (b) the lower and upper parabolas correspond to the dispersion laws of electrons near the stopping point and near the top of the barrier, respectively; required momentum scattering is indicated by a green horizontal dashed arrow.

It is worth noting that in the proposed simplified picture, the "resonant" optical transition is direct (vertical) in $k$-space (Fig. 1b), but it is indirect in the real, $r$-space, that is, under the transition, the electron "shifts" in the real space from the stopping point to the top of the barrier (Fig. 1a). In reality, the electron states are delocalized both in real space and in $k$-space, so the simple picture of electron optical transitions between the states with well-defined dispersion laws is valid only for sufficiently smooth barriers, for which the momentum uncertainty is relatively small. From the Heisenberg uncertainty relation, it is possible to estimate the half-width of the spectral maximum $\Delta E \sim (\Delta p)^2 / 2m^*$, where $m^*$ is the electron effective mass, $\Delta p$ is the quasimomentum acquired by the electron under transfer through QPC. Taking the uncertainty of the coordinate equal to the half-width of the potential barrier at the Fermi level $\Delta x \sim 30$ nm, we obtain $\Delta E \sim 0.5$ meV, which agrees with the results of the numerical calculations [13].

In order to carry out a qualitative analysis of various factors that contribute to the formation of the spectral peak in the photoconductance, we have calculated PST spectra as the product of the optical transition probability $W$ and the electron transfer probability $D$ through the potential barrier in the final state. It should be noted that the idea to explore a similar product in order to explain the resonant-like spectral maximum in PST was proposed, although not fully realized in ref. [7]. Here the probability of the optical transitions from the initial electron state with energy $E_F$ to the final state with energy $E_F + \hbar\omega_0$ was calculated according to the first order perturbation theory, using the golden Fermi rule $W = 2\pi/\hbar \cdot |\langle\psi_f|H'|\psi_i\rangle|^2 \rho(E_F + \hbar\omega_0)$, where $\rho(E)$ is the one-dimensional density of states. The wave functions of the initial $\psi_i$ and final $\psi_f$ states were taken from the solution of the problem of electron transfer through a smooth Eckart barrier with characteristic width $d$: $U(x) = U_0/\cosh^2(x/d)$ [15]. The Hamiltonian $H'$ of the interaction between electron and electromagnetic radiation was taken from [16]. The electron transmission coefficient for the final state $D(E_F + \hbar\omega_0)$ was also taken from [15]. The calculations were done for the intensity of terahertz radiation independent from $\hbar\omega$ and equal to 200 mW/cm$^2$.

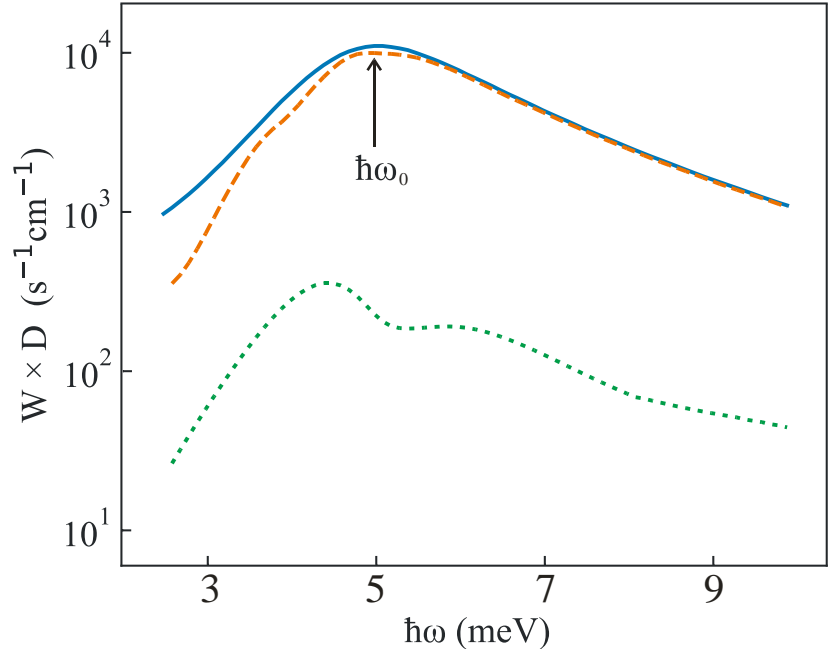

Fig. 2. Calculated PST spectra of a QPC for $d = 100$ nm, $U_0 = 30$ meV, $E_F = 25$ meV and for the matrix element of optical transitions integrated over various space regions $L$ with respect to the characteristic barrier width $d$: $L=100d$ (blue solid line); $L=4d$ (orange dashed line); $d$ (green dotted line). The "resonant" photon energy $\hbar\omega_0 = U_0 - E_F = 5$ meV is indicated by the arrow.

Shown in Fig. 2 are the spectra of PST ~ $W \times D$ calculated for $d = 100$ nm, $U_0 = 30$ meV, $E_F = 25$ meV and for the matrix element of optical transitions integrated over the space regions from $-L$ to $+L$ around the center of the barrier, for various $L$. The spectral region is restricted by the photon energies $\hbar\omega > (U_0 - E_F)/2$, because, for lower photon energies multi-photon processes, which are not accounted for in the present perturbation theory calculations, yield significant contribution to the photon-stimulated transport [14]. It is seen that all spectra contain the leading peak centered approximately at the photon energy $\hbar\omega_0 \approx U_0 - E_F = 5$ meV corresponding to optical transitions from the Fermi level to the top of the barrier. At $\hbar\omega < \hbar\omega_0$, the increase of the PST with increasing photon energy is due to the increase in $D$; at $\hbar\omega > \hbar\omega_0$, the decrease of the PST is due to the decrease in the probability of optical transition $W$, while $D$ saturates to an approximately constant value; thus, the resonant-like peak is formed at $\hbar\omega \approx \hbar\omega_0$. These considerations are in line with a qualitative explanation of the resonant-like peak in PST of a QPC proposed earlier in [7].

The comparison of the spectra for various $L$ shows that the main contribution to the resonant-like leading peak comes from the integration of the matrix element over the region with $L = 4d$; in this region the electron wave function is modified by the potential barrier so that the optical transitions become possible. It is seen that the decrease in $L$ down to $L = d$ leads to drastic decrease in the peak amplitude, while for larger $L > 4d$, amplitude and shape of the peak saturate with further increase in $L$. It should be noted that for the chosen energies and barrier parameters, the stopping points lie within $L \sim d$. This fact restricts a simple assumption that the optical transitions at the stopping points yield the major contribution to the PST.

Relatively weak additional shoulders and extremes, which are superimposed on the leading peak for small integration regions $L = d$ and $L = 4d$, are the "side lobes" arising due to the effect of finite integration window. Another artifact consisted in the noise-like undulations in the PST spectra due to different phase of the integrand $\psi_f H' \psi_i$ at the boundaries of the integration range for different $\hbar\omega$; these undulations were suppressed by averaging over the phase.

Fig. 3 shows the PST spectra calculated for various Fermi levels $E_F$. It is seen that for all values of $E_F$, the spectral position of the leading peak corresponds to the optical transitions from the Fermi level to the top of the potential barrier, in accordance with the results of the numerical calculations performed in [13] and with the qualitative explanation of PST proposed here (Fig. 1). With increasing $E_F$, the peak broadens and becomes less distinct. This fact confirms the abovementioned considerations based on the uncertainty relation. It is also seen that the slope of the high-energy tail of the peak weakly depends on the Fermi level. It should be noted that, along with the drop in the optical absorption due to restrictions imposed by the momentum conservation law, a decrease in the density of states in a quasi-one-dimensional channel with an increase in the kinetic energy of electrons contributes to the drop in the photoconductance with an increase in $\hbar\omega$. This factor was noted in [13], but, apparently, it is not dominant, since the density of states decreases with energy by a power-law $\rho(E) \sim E^{-1/2}$, while the photoconductance decreases with increasing photon energy much faster (exponentially, judging by the results of the calculations [13]). Also, in the present calculations of the optical transition probability $W$, we used the final states density calculated for kinetic energy $E = E_F + \hbar\omega_0$ away from the barrier region, while, near the top of the barrier, smaller kinetic energy and, thus enhanced PST should be expected. This factor should further sharpen the resonant-like peak corresponding to PST with transitions to the top of the barrier. Presumably, this is the reason for the sharper spectral peaks resulting from numerical calculations, as compared to those obtained here on the basis of the perturbation theory.

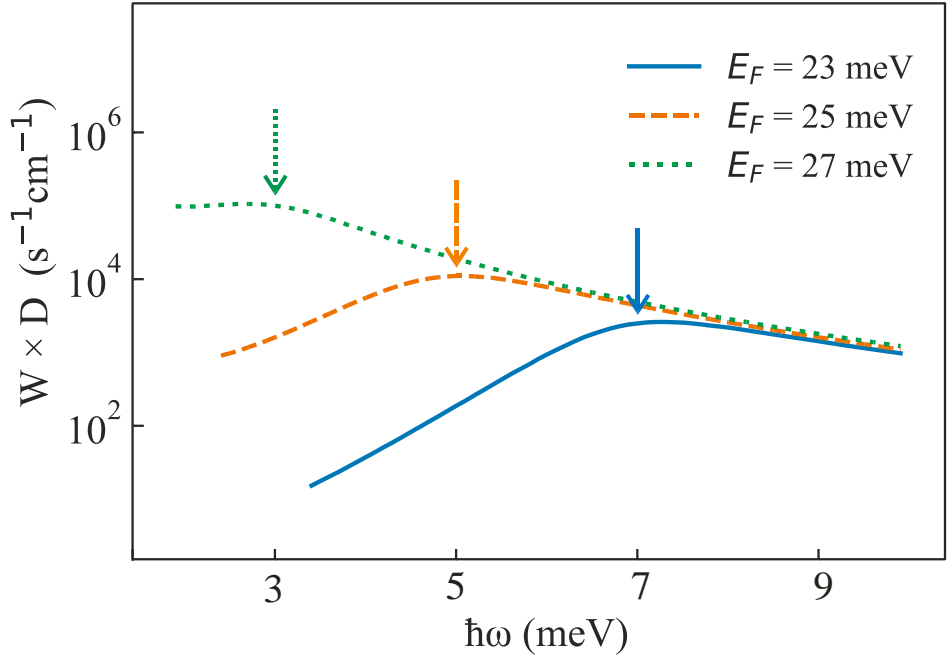

Fig. 3. PST spectra of a QPC calculated for $d = 100$ nm, $U_0 = 30$ meV and various positions of the Fermi level: $E_F = 23$ meV (blue solid line); 25 meV (dashed orange line); 27 meV (green dotted line). For each spectrum, the respective "resonant" photon energy $\hbar\omega_0 = U_0 - E_F$ is indicated by the arrow.

Thus, considerations about the momentum conservation upon absorption of photons provide a qualitative explanation for the nonmonotonic, resonant-like spectral dependence of the photoconductance of a QPC with a smooth potential barrier, which was observed experimentally [12] and obtained by numerical calculations [13]. Within the proposed physical picture, the widths of the spectral maxima are estimated, and the evolution of the shape of the spectra with a change in the position of the Fermi level is qualitatively explained. We have calculated the spectra of the PST on the basis of the perturbation theory. The results also predict the spectral maxima in PST for optical transitions from the Fermi level to the top of the potential barrier, in agreement with the numerical calculations [13]. Similar resonant spectral features can be expected in the photoconductance of other nanostructures, as well as in the photoemission of electrons from metals and heavily doped *n*-type semiconductors into vacuum.

This work was supported by the Russian Foundation for Basic Research, Grant No. 20-02-00355.

**References**

[1] A. H. Dayem, R. J. Martin, Quantum interaction of microwave radiation with tunneling between superconductors, Phys. Rev. Lett. **8**, 246 (1962).

[2] P. K. Tien, J. P. Gordon, Multiphoton process observed in the interaction of microwave fields with the tunneling between superconductor films, Phys. Rev. **129**, 647 (1963).

[3] M. Büttiker, R. Landauer, Traversal time for tunneling, Phys. Rev. Lett. **49**, 1739 (1982).

[4] D. D. Coon, H. C. Liu, Time-dependent quantum-well and finite-superlattice tunneling, J. Appl. Phys. **58**, 2230 (1985).

[5] M. Grifoni, P. Hänggi, Driven quantum tunneling, Phys. Rep. **304**, 229 (1998).

[6] G. Platero, R. Aguado, Photon-assisted transport in semiconductor nanostructures, Phys. Rep. **395**, 1 (2004).

[7] J.-Y. Ge, J. Z. H. Zhang, Quantum mechanical tunneling through a time-dependent barrier, J. Chem. Phys. **105**, 8628 (1996).

[8] K. Yakubo, S. Feng, Q. Hu, Simulation studies of photon-assisted quantum transport, Phys. Rev. B. **54**, 7987 (1996).

[9] O. A. Tkachenko, V. A. Tkachenko, D. G. Baksheyev, H. Nejoh, Localization of tunneling electron in a potential barrier with alternating height, Proc. 5th Int. Symp. Foundations of Quantum Mechanics in the Light of New Technology (ISQM-Tokyo '95), Japan, August 21-24, 1995 / ed. by K. Fujikawa, Y..A. Ono., New York: Elsevier/North Holland, 1996, P. 207.

[10] S. Morina, O. V. Kibis, A. A. Pervishko, I. A. Shelykh, Transport properties of a two-dimensional electron gas dressed by light, Phys. Rev. B **91**, 155312 (2015).

[11] V. M. Kovalev, W.-K. Tse, M. V. Fistul, I. G. Savenko, Valley Hall transport of photon-dressed quasiparticles in two-dimensional Dirac semiconductors, New J. Phys. **20**, 083007 (2018).

[12] M. Otteneder, Z. D. Kvon, O. A. Tkachenko, V. A. Tkachenko, A. S. Jaroshevich, E. E. Rodyakina, A. V. Latyshev, S. D. Ganichev, Giant terahertz photoconductance of quantum point contacts in the tunneling regime, Phys. Rev. Applied. **10**, 014015 (2018).

[13] O. A. Tkachenko, V. A. Tkachenko, D. G. Baksheev, Z. D. Kvon, Steps of the giant terahertz photoconductance of a tunneling point contact, JETP Lett. **108**, 396 (2018).

[14] V. A. Tkachenko, Z. D. Kvon, O. A. Tkachenko, A. S. Yaroshevich, E. E. Rodyakina, D. G. Baksheev, A. V. Latyshev, Photon-Stimulated Transport in a Quantum Point Contact, JETP Lett. **113**, 331 (2021).

[15] L. D. Landau, E. M. Lifshitz, Quantum Mechanics: Non-Relativistic Theory, Vol. 3, P. 80 (3rd ed.), Pergamon Press, 1977.

[16] A. Anselm, Introduction to Semiconductor Theory, Mir Publishers, Moscow, 1981, P. 406.